\documentclass[aps,prb,twocolumn,superscriptaddress]{revtex4-2}
\usepackage{graphicx,amsmath,amssymb,bm,booktabs}

\graphicspath{{./}{../figures/}}

\begin{document}

\title{Shift current as a stacking spectroscopy of hexagonal boron
nitride multilayers}

\author{Sanber Vizcaya}
\affiliation{Grupo de Simulaciones, Departamento de F\'isica,
Universidad T\'ecnica Federico Santa Mar\'ia, Avenida Espa\~na 1680,
Valpara\'iso 2390123, Chile}
\author{Juan M. Florez}
\affiliation{Facultad de Ingenier\'ia y Arquitectura, Universidad
Central de Chile, Avenida Santa Isabel 1186, Santiago 8330601, Chile}
\author{Eric Su\'arez Morell}
\email{eric.suarez@usm.cl}
\affiliation{Grupo de Simulaciones, Departamento de F\'isica,
Universidad T\'ecnica Federico Santa Mar\'ia, Avenida Espa\~na 1680,
Valpara\'iso 2390123, Chile}

\begin{abstract}
Out-of-plane polarization can be tuned by sliding polar monolayers in
van der Waals materials; the bulk photovoltaic effect is emerging as a
natural optical probe, provided the photocurrent tracks the
polarization. We test that assumption by computing the polarization
and the shift-current tensor from first principles across the stacking
landscape of bi-, tri-, and tetralayer hexagonal boron nitride. The
two quantities share one robust property: the vertical response
reverses with the polarization while the in-plane one does not. Beyond
this rule, the correspondence must be read with caution: stackings
with identical polarization carry vertical monochromatic photocurrents
of opposite sign; the spectra reorganize and even reverse sign at some
photon energies along a constant-polarization plateau; and the
strongest vertical peak of the entire pathway sits where the
polarization crosses zero. The reason is structural: the polarization
counts polar interfaces, while the shift current reads their
positions. Broadband illumination reads the domain sign; monochromatic
photocurrent maps can read the stacking registry. In a sliding
ferroelectric, the bulk photovoltaic effect can operate as a stacking
spectroscopy.
\end{abstract}

\maketitle

\section{Introduction}\label{sec:intro}

The natural optical probe of a polar crystal is the bulk photovoltaic effect (BPVE): a dc current generated by uniform illumination of a homogeneous, junction-free material. Its dominant intrinsic mechanism in clean semiconductors is the shift current \cite{vonBaltz1981,Sipe2000,Young2012,Cook2017}: in a noncentrosymmetric crystal the electron clouds of valence and conduction states at the same $\bm{k}$ are centered at different positions, so every absorbed photon displaces charge by a definite vector, the shift vector, and those displacements add up to a macroscopic current.

For sliding ferroelectrics, layered polar crystals whose polarization is set by the lateral registry between rigid layers \cite{Li2017,Yasuda2021,Stern2021,Wu2021}, the connection with the polarization seems almost too good: both quantities are wave-function (Berry-phase or quantum-geometric) properties, and symmetry analysis shows that the out-of-plane photocurrent switches with the polarization while the in-plane one does not; this is the ``non-synchronous'' BPVE \cite{Xiao2022,Kaner2022}. It is tempting to conclude that the photocurrent reads the polarization, and parts of the experimental literature lean on precisely that reading \cite{NbOBr2}.

Experiments as a whole, however, are contradictory. In CuInP$_2$S$_6$ the BPVE photocurrent reverses sign at fixed polarization direction, its authors noting that ``the exact correlation between the photocurrent and spontaneous polarization remains unclear'' \cite{Bai2024}; computed shift conductivities in the same material underestimate a measured thickness enhancement by two orders of magnitude \cite{Delodovici2025}; the thickness dependence of the BPVE in 3R-MoS$_2$ escapes conventional transport models \cite{Xu2025}; and in NbOBr$_2$ a direct proportionality between shift vector and polarization difference has been invoked with apparent success \cite{NbOBr2}. Measurements alone cannot reconcile these claims, because experiments compare only a handful of discrete samples. What is needed is a setting where polarization and shift current can be computed exactly, side by side, while the ferroelectric order parameter is tuned continuously. The comparison itself has to be drawn with care: the polarization (for a given direction) is a single number, while the shift current is a spectrum. Any claimed relation between them is therefore a statement about a projection, either the frequency-integrated response or the response at a fixed photon energy, and the two projections need not behave alike.

Multilayer hexagonal boron nitride (hBN) addresses this need: light elements, a wide band gap, and quantitatively parametrized interlayer couplings \cite{Cortes2023,Vizcaya2025} make it the cleanest sliding ferroelectric.

The mechanism is geometric. In the AB and BA registries, a boron atom of one layer sits directly above a nitrogen atom of the other; that pairing dictates the interlayer charge balance. Sliding one layer by a single bond length converts AB into BA and reverses the dipole. As the number of layers increases, the space of stacking configurations grows into a landscape of polytypes and polarization values \cite{Cortes2023,Vizcaya2025}. 

Some sliding pathways contain extended polarization plateaus, in which $P_z$ remains essentially unchanged while the microscopic stacking configuration continues to evolve \cite{Vizcaya2025}. These plateaus provide a stringent test of the relationship between polarization and shift current.  Whether the photocurrent can see what the polarization cannot is the question we answer.

This article follows both quantities across the stacking landscape of hBN. We compute the polarization and shift-current response from first principles, continuously along the sliding pathway of the bilayer (Sec.~\ref{sec:bilayer}), over all nine registry stackings of the trilayer (Sec.~\ref{sec:trilayer}), and along a low-barrier sliding pathway of the tetralayer that contains a polarization plateau (Sec.~\ref{sec:tetralayer}). Two complementary methods make this analysis possible (Sec.~\ref{sec:model}): Wannier-based Hamiltonians, one per stacking, that provide the quantitative shift-current results; and an analytic tight-binding Hamiltonian that supplies continuous polarization pathways and guides the exploration.

We found that the polarization and the shift current share exactly one robust property, the mirror-dictated switching rule; beyond it they decouple, and the decoupling has a microscopic origin (Sec.~\ref{sec:shiftvector}): the polarization is an occupied-state, whole-crystal quantity that merely counts polar interfaces, while the shift current is a transition-resolved $\pi$-manifold quantity that weighs each optical excitation by its vertical displacement and therefore reads the location of the interfaces. 
What emerges is a set of rules (Sec.~\ref{sec:discussion}): monochromatic photocurrent maps can read the registry, broadband illumination reads the domain sign, and the in-plane channel provides a sliding-invariant normalization.

\section{Theoretical and computational methods}\label{sec:model}

\subsection{The shift-current tensor}

For linearly polarized light the shift-current conductivity is
\cite{Sipe2000,Aversa1995}
\begin{equation}
\sigma^{a;bc}(\omega) = \frac{\pi e^3}{\hbar^2}\!\sum_{n,m}\!\int\!
\frac{d^2k}{(2\pi)^2}\, f_{nm}\,
\mathrm{Im}\!\left[ r^{b}_{mn} r^{c}_{nm;a} \right]
\delta(\omega_{mn}-\omega),
\label{eq:sipe}
\end{equation}
where $a$ is the direction of the current and $b$, $c$ are the
Cartesian components of the light polarization (the expression is
symmetrized over $b$ and $c$). The band indices $n$, $m$ run over the
Bloch states at each $\bm{k}$; $f_{nm} = f_n - f_m$ is their occupation
difference, $\hbar\omega_{nm} = E_n - E_m$ their energy difference, and
$v^{a}_{nm} = \langle n\bm{k}|\hat{v}^{a}|m\bm{k}\rangle$ the matrix
element of the velocity operator. For $n \neq m$, $r^{a}_{nm} =
v^{a}_{nm}/(i\omega_{nm})$ is the interband (transition) dipole matrix
element, and $r^{c}_{nm;a}$ is its generalized (covariant) derivative
with respect to $k_a$. Since a naive $\bm{k}$ derivative of $r_{nm}$ is
gauge dependent, the covariant derivative is evaluated through the
sum-over-states identity
\begin{align}
r^{c}_{nm;a} = \frac{i}{\omega_{nm}}\Bigl[&
\frac{v^{c}_{nm}\Delta^{a}_{nm} + v^{a}_{nm}\Delta^{c}_{nm}}{\omega_{nm}}
- h^{ac}_{nm} \nonumber\\
&+ \sum_{p\neq n,m}\Bigl(
\frac{v^{c}_{np}v^{a}_{pm}}{\omega_{pm}} -
\frac{v^{a}_{np}v^{c}_{pm}}{\omega_{np}} \Bigr)\Bigr],
\label{eq:gencov}
\end{align}
with $\Delta^{a}_{nm} = v^a_{nn} - v^a_{mm}$ the difference of band
velocities and $h^{ac}_{nm}$ the matrix element of the second derivative of the Hamiltonian, $h^{ac}_{nm} = \langle n\bm{k}|\partial^2 \hat{H}/\partial k_a \partial k_c|m\bm{k}\rangle$; the sum over intermediate states $p$
encodes the virtual channels whose role we quantify in
Sec.~\ref{sec:bands}. Throughout the article $\omega$ denotes a
frequency (an energy difference divided by $\hbar$), while the symbol
$w$ is reserved for an optical weight: the transition probability
$|r^{b}_{cv}|^2$ of a valence--conduction pair or its zone integral
(Sec.~\ref{sec:shiftvector}).

The quantity with direct physical meaning is the shift vector,
$R^{a}_{nm} = \partial_{k_a}\!\arg r^{b}_{nm} + A^{a}_{nn} - A^{a}_{mm}$,
where $A^{a}_{nn} = \langle u_n | i\partial_{k_a} u_n\rangle$ is the
intraband Berry connection and $b$ is the Cartesian index of the light
polarization. It is the
gauge-invariant displacement of the electron cloud per absorbed photon.
Its first term is a phase derivative: the phase can wind around the
Brillouin zone, so the average of $R$ is defined only up to an integer
number of lattice vectors, the analog of the polarization quantum. 

A slab removes this subtlety for
the vertical component. With no $k_z$ there is no phase derivative
along $z$, hence no winding, and the out-of-plane shift vector reduces
exactly to
\begin{equation}
R^{z}_{cv}(\bm{k}) = z_{cc}(\bm{k}) - z_{vv}(\bm{k}),
\label{eq:rz}
\end{equation}
the vertical distance between the charge centroids of the two states.
The vertical response weighs this jump, for every valence--conduction
pair $(v,c)$ and every $\bm{k}$, by the absorption strength of the
transition,
\begin{equation}
\sigma^{z;bb}(\omega) \propto \sum_{cv}\int\!\frac{d^2k}{(2\pi)^2}\,
R^{z}_{cv}(\bm{k})\,|r^{b}_{cv}(\bm{k})|^{2}\,\delta(\omega_{cv}-\omega),
\label{eq:rzweight}
\end{equation}
where $|r^{b}_{cv}|^{2}$ is the optical transition probability of the
pair for light polarized along $b$; averaged over the in-plane
polarization it is the weight $w_{cv}$ of Sec.~\ref{sec:shiftvector}.

Throughout this article the light is in-plane polarized.
The vertical photocurrent is then carried by $\sigma^{zxx}$ and
$\sigma^{zyy}$ (from here on we drop the semicolon of
Eq.~\eqref{eq:sipe}, $\sigma^{abc} \equiv \sigma^{a;bc}$), which
coincide at every $C_{3v}$ registry; along
continuous pathways we also use their rotation-invariant average
$\sigma_{\rm vert} = \tfrac12(\sigma^{zxx}+\sigma^{zyy})$. The in-plane
channel is $\sigma^{yyy}$. The remaining vertical component allowed at
$C_{3v}$ stackings, $\sigma^{zzz}$, absorbs only through an
out-of-plane field and would require oblique incidence; we do not consider it.
In an isolated flake the vertical response appears as a photo-induced
interlayer voltage \cite{Xiao2022}; between vertical contacts it is
read directly as a photocurrent.

Conductivities are quoted as sheet values, spin factor included. Because
the system is a slab, the integral in Eq.~\eqref{eq:sipe} runs over the
two-dimensional Brillouin zone and $\sigma^{a;bc}$ relates the
photocurrent per unit width, $J^{a}_{\rm 2D} = \sigma^{a;bc}E_bE_c$ (in
A/m), to the field amplitudes of the light. Its unit is therefore
A$\cdot$m/V$^2$, which we write as $\mu$A$\cdot$nm/V$^2$. It equals the
bulk conductivity of the three-dimensional literature ($\mu$A/V$^2$)
integrated over the thickness of the slab, $\sigma_{\rm 2D} =
\int\sigma_{\rm 3D}\,dz$, so a bulk value is recovered by dividing by an
effective thickness (we use this conversion in Sec.~\ref{sec:bands}).
Experimentally, $\sigma_{\rm 2D}$ is the measured short-circuit
photocurrent divided by the width of the illuminated channel and by the
square of the field amplitude, the latter obtained from the incident
intensity; no thickness enters, which is the reason to prefer sheet
values for atomically thin samples. Frequency integrals of $\sigma$ are
given in $\mu$A$\,$nm/V$^2\,$eV.

\subsection{Wannier-based Hamiltonians}\label{sec:wannier}

Every quantitative shift-current statement in this article comes from density-functional theory. At each stacking (46 in total: 12 bilayer slidings, 9 trilayers, 25 tetralayer slidings) we perform a plane-wave calculation (\textsc{Quantum ESPRESSO} \cite{Giannozzi2009}, PBE with Grimme-D3, dipole-corrected slabs). The plane-wave kinetic-energy cutoff was set to $E_{\mathrm{cut}}^{\mathrm{wf}} = 80~\mathrm{Ry}$, with a charge-density cutoff of
$E_{\mathrm{cut}}^{\rho} = 320~\mathrm{Ry}$. Brillouin-zone integrations were performed using a
$12 \times 12 \times 1$ Monkhorst--Pack mesh, with a uniform
$24 \times 24 \times 1$ grid for the non-self-consistent step that feeds the Wannier construction. A vacuum region of at least $18~\text{\AA}$ was used along the $z$ direction (cells of $c = 25$~\AA{} for bilayer and trilayer, $35$~\AA{} for the tetralayer) to avoid interactions between periodically repeated images.
The electronic self-consistency threshold was set to
$10^{-10}~\mathrm{Ry}$. 

We are aware that not every stacking in this set is energetically stable. The unstable registries serve as reference configurations: the aim is the behavior of both
observables across the whole registry landscape. 

Every value of polarization comes directly from the self-consistent calculation, as the slab dipole per unit area evaluated by the dipole correction, with no Wannier construction involved. We then
Wannierize the $\pi$ manifold ($2N$ bands for $N$ layers) with projectability-disentangled Wannier functions \cite{Qiao2023,Pizzi2020}, reproducing the DFT $\pi$ valence at K to sub-meV precision and the $\pi^{*}$ doublet to a few tens of meV.

We use the $\pi$-orbital subspace because it hosts every optical
resonance of the analysis window and its Wannierization is
unambiguous. Still, we verified for five
structures, including at least one for each multilayer type, that the
results obtained when the full $\sigma$ valence is included did not
differ significantly.
Every sign, onset, and zero crossing of the spectra survives; the
integrated responses move by a few percent, at most amplified, never
erased or reversed in sign.

The Wannier-based (TBW) Hamiltonian and the full position-operator matrix
elements $\langle \bm{0}n|\hat{\bm r}|\bm{R}m\rangle$ supply every
quantity entering Eqs.~\eqref{eq:sipe}--\eqref{eq:rz}
\cite{Ibanez2018}. 

The same position-operator matrix elements yield every band-resolved quantity in this article. At each $\bm{k}$, rotating the Wannier position operator to the eigenstate basis gives the vertical
band centroid $z_{nn}(\bm{k}) = \langle u_{n\bm{k}}|\hat{z}|u_{n\bm{k}} \rangle$, the charge center of the Bloch state entering Eq.~\eqref{eq:rz}. In a slab this diagonal element is a genuine expectation value, free of the winding ambiguity discussed above, so the band centroids and their zone averages $\langle z\rangle_n$ are well defined. The $\pi$-manifold polarization, $P^{\pi}_z$ (written $P^{\pi}$ for short), is reconstructed from the occupied centroids, with the compensating ionic charge placed at the atomic $z$ of nitrogen, fixed along a rigid pathway.

The position operator of a periodic cell is defined through the phases
$e^{-i2\pi z/c}$ \cite{Resta1998}, and the finite differences that
build its Wannier matrix elements are not invariant under a
translation of the slab within the cell \cite{Lihm2026}. Centring it at
$z = c/2$, the natural choice for a dipole-corrected calculation,
places the atoms exactly on the branch cut of those phases; in
our calculations this produced wrapped Wannier centres, mixed-branch
$z$ matrix elements, and quantitatively unreliable vertical responses. All slabs in this work
are therefore recentred at $z = 0$.

\subsection{The tight-binding Hamiltonian}\label{sec:tb}

The exploration that led to this article was guided by an analytic $p_z$ tight-binding Hamiltonian \cite{Cortes2023,Vizcaya2025}: on-site energies $\pm\Delta$, nearest-neighbor intralayer hopping $t_\parallel$, and species-resolved exponential interlayer hoppings $t_{XY}(r) = t^{0}_{XY} e^{-\alpha(r-d_0)}$ between atoms of species $X$ and $Y$ at distance $r$, with $\alpha$ the decay constant and $d_0$ the reference interlayer distance of the parametrization, whose B--B/N--N/B--N asymmetry is the microscopic origin of everything polar in this problem. For this model, the polarization follows the same sliding dependence as the first-principles result, with its absolute magnitude related by a multiplicative factor \cite{Cortes2023,Vizcaya2025}.

For polarization calculations, the TB $P_z$ tracks the first-principles dipole along entire pathways, so we use it for continuous curves $P_z(s)$. Here $s$ is the sliding coordinate used throughout the article: the rigid in-plane displacement of one layer in units of the cell diagonal $\bm{a}_1+\bm{a}_2$, so that $s = 0$ at AA, $s = 1/3$ at AB and $s = 2/3$ at BA (Sec.~\ref{sec:sweep}). The TB Hamiltonian also respects every group-theoretic constraint exactly, so it reliably predicts selection rules and switching behavior.

However, for the magnitude of the vertical response, the model fails. Already at AB the two TB calibrations, to the bands or to the polarization, differ by a factor of thirty in the frequency-integrated vertical response ($0.66$ and $0.02~\mu$A$\,$nm/V$^2\,$eV, against the first-principles $0.70$), and their peak values ($2.71$ and $0.19~\mu$A$\cdot$nm/V$^2$) bracket the first-principles peak of $1.51$ without reproducing its spectral structure. The two calibrations differ in the single interlayer parameter $t^{0}_{BN}$, which cannot fit the interlayer band splittings and the interlayer charge transfer at the same time. Most dramatically, the TB predicts a near-vanishing vertical response for the trilayer ABC where first principles give the largest (Sec.~\ref{sec:trilayer}). The reason is that the vertical magnitude is built by remote-band virtual channels \cite{Chen2024,PerezRiffo2026} (Sec.~\ref{sec:bands}) that no minimal $p_z$ parametrization captures. 

The failure is not necessarily due to the size of the Hilbert space. The $\pi$ TBW Hamiltonian spans the same four states. Its hoppings and
position matrix elements, however, are exact projections of the DFT
problem, and they carry the imprint of the remote bands on the $\pi$
block; the local parametrization carries none of it. 
Disentangling the remote-band contribution (Sec.~\ref{sec:wannier}), we found that the occupied half contributes essentially nothing, so the missing weight resides in the conduction states. The conduction half is anything but negligible: it sets the absolute scale of the vertical response (Sec.~\ref{sec:bands}). A quantitative treatment of those channels lies beyond any local model, and we leave it for a dedicated study. 


\subsection{Symmetry of the stacking landscape}\label{sec:symm}

Registry stackings fall into two point groups (Table~\ref{tab:symm}; general framework in Ref.~\cite{Ji2023}). Nonpolar stackings, those whose
interfaces cancel, belong to $D_{3h}$, whose horizontal mirror $m_z$ forbids both $P_z$ and any vertical response, but not the in-plane one: breaking inversion within the plane (B versus N) suffices for an in-plane BPVE, as the monolayer already shows. Every polar registry stacking, for any layer number, belongs to $C_{3v}$.

Reflecting a stacking through $m_z$ yields its
opposite-polarization partner (Table~\ref{tab:symm}), up to an
in-plane translation that realigns the registries. Under that map every tensor component acquires one factor of $-1$ per $z$ index. The consequence is the generalized non-synchronous switching rule \cite{Xiao2022}:
components with an odd number of $z$ indices, like $P_z$ itself and $\sigma^{zxx}$, reverse with the domain; components with an even number, like $\sigma^{yyy}$, do not.  That is why the vertical maps below are antisymmetric under $s \leftrightarrow 1-s$ while the in-plane ones are symmetric. 

\begin{table}[tb]
\caption{\label{tab:symm}Registry stackings (mirror pairs $X/Y$ listed
together), point groups (verified with \textsc{spglib}), and raw
TB polarization. $D_{3h}$ stackings are self-mirror; the $C_{3v}$
pairs are $m_z$ partners, with ABB$\leftrightarrow$AAC and
AAB$\leftrightarrow$ACC crossed ($^\ast$). All groups allow the
$\sigma^{yyy}$ family; $C_{3v}$ additionally allows
$\sigma^{zxx}\!=\!\sigma^{zyy}$ and $\sigma^{zzz}$.}
\begin{tabular}{lcc}
\toprule
Stacking & Group & $P_z$ (pC/m)\\
\midrule
AB / BA & $C_{3v}$ & $\pm 0.716$\\
AA, ABA, ACA, AAA, ABBA & $D_{3h}$ & $0$\\
ABC / ACB & $C_{3v}$ & $\pm 1.432$\\
ABB / AAC$^\ast$ & $C_{3v}$ & $\pm 0.72$\\
AAB / ACC$^\ast$ & $C_{3v}$ & $\pm 0.72$\\
ABAC / ABCB & $C_{3v}$ & $\mp 0.716$\\
ABCA / ACBA & $C_{3v}$ & $\pm 2.148$\\
ABAB / CBCB & $C_{3v}$ & $\pm 0.716$\\
\bottomrule
\end{tabular}
\end{table}

\section{Polarization and shift current in the bilayer}\label{sec:bilayer}

\subsection{Response along the sliding pathway}\label{sec:sweep}

\begin{figure}[tb]
\includegraphics[width=\columnwidth]{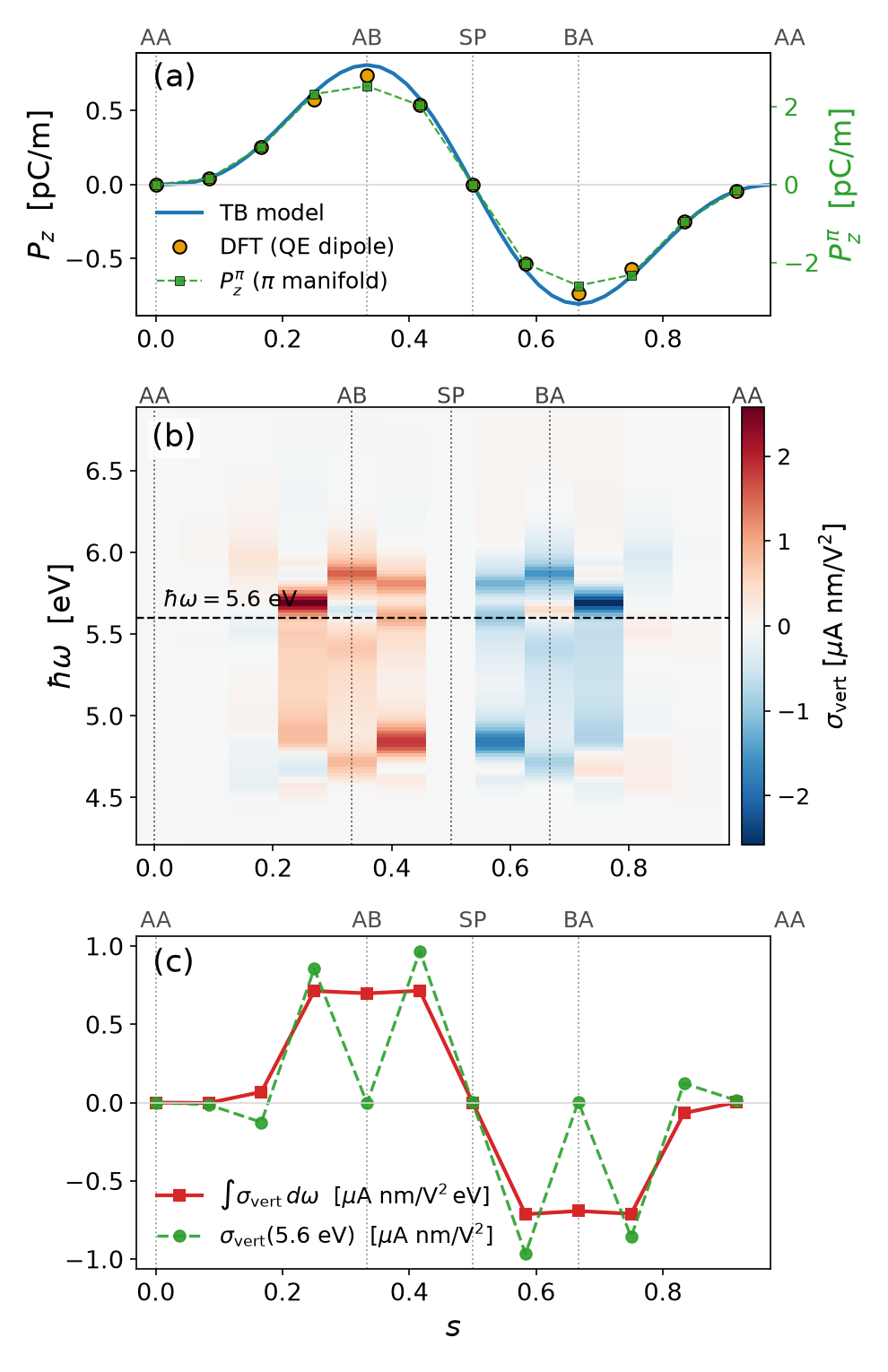}
\caption{\label{fig:sweep}First-principles sweep of the bilayer diagonal
pathway AA$\to$AB$\to$BA$\to$AA (twelve slidings, rigid layers,
$d=3.429$~\AA). The sliding parameter $s$ displaces the top layer by
$\bm{\tau} = s\,(\bm{a}_1+\bm{a}_2)$ along the cell diagonal (a)~Dipole-corrected DFT polarization (points) versus the
TB Hamiltonian evaluated continuously along the pathway (line). Green
squares, right axis: the $\pi$-manifold polarization $P_z^{\pi}$
(Sec.~\ref{sec:shiftvector}); with the right axis scaled by the
screening ratio, it falls on the same curve. (b)~Full spectral map of
the vertical response $\sigma_{\rm vert}(\omega,s)$. (c) Under
broadband illumination the integrated response is a
step function of the domain sign; the
monochromatic response at $5.6$~eV vanishes at every high-symmetry
stacking and peaks only at intermediate registries.}
\end{figure}


The bilayer is the minimal system, with one interface and one polar degree of freedom. We therefore use it to establish the mechanisms
underlying our conclusions, which are then tested in the trilayer and tetralayer of Secs.~\ref{sec:trilayer} and~\ref{sec:tetralayer}.

In Fig.~\ref{fig:sweep} we show the out-of-plane polarization and the vertical shift-current response along the diagonal sliding pathway $AA\to AB\to BA\to AA$, parametrized by the displacement of the top layer, $\bm{\tau}=s\,(\bm{a}_1+\bm{a}_2)$ with $s\in[0,1]$. Panel (a) shows the polarization, which is antisymmetric about the saddle-point stacking $SP$ ($s=\tfrac12$), reaches opposite extrema at the polar registries $AB/BA$, and vanishes at the eclipsed $AA$ stacking. This stacking dependence is well established from first-principles calculations of bilayer and multilayer hBN \cite{Li2017,Cortes2023,Vizcaya2025} and has also been observed experimentally in twisted-hBN domain structures \cite{Woods2021}. Our sweep reproduces this behavior, with the twelve first-principles dipoles following the TB curve. Panel (b) shows the vertical shift-current spectrum along the same pathway. The $AA$ and $SP$ columns vanish by symmetry, and the map is antisymmetric under $s\leftrightarrow1-s$. Beyond these symmetry-dictated features, the map contains the first unexpected result: the strong response lobes appear exclusively at intermediate, low symmetry slidings.

At the polar AB/BA registries the spectral weight is redistributed.
Instead of one dominant peak, the AB spectrum shows three maxima, at
$4.7$, $5.4$ and $5.9$~eV, and between the last two it reverses sign
within a narrow frequency interval, $5.58$--$5.68$~eV, centered at
$5.6$~eV; we refer to it below as the reversal interval. The
monochromatic response at $5.6$~eV [panel (c), dashed line] therefore
reads zero at every high-symmetry stacking, polar or not, and peaks
($\pm 0.97~\mu$A$\,$nm/V$^2$) only at intermediate registries.
Where the polarization is largest, the monochromatic photocurrent
vanishes. Panel (c) also shows the opposite face: the
frequency-integrated response is a clean $\pm
0.71~\mu$A$\,$nm/V$^2\,$eV step function of the domain sign, essentially
constant on each side while the spectrum underneath reorganizes
completely.

The zeros at AB/BA mark that sign reversal, and the reversal is forced by three properties of the $\pi$ manifold: a trace sum
rule guarantees that both signs of vertical displacement are present at
every $\bm{k}$; the optical weight resides in two layer-conserving
channels whose displacements are opposite; and these two channels are
split by $0.22$~eV, precisely where the van Hove singularity of M
concentrates the spectral weight. 

The first two properties can be read from the band centroids of Fig.~\ref{fig:shiftvector}(a): the four centroids compensate one another, and the conduction doublet lies
closer to the slab center than the valence doublet, so both bright
transitions displace the electron toward the center, upward from the
bottom layer and downward from the top one; hence the opposite signs.
Within the $0.22$~eV splitting of the two bright channels their
opposite contributions compete, and where they balance the spectrum
crosses zero. At AB this happens twice, at $5.58$ and $5.68$~eV, which
leaves a narrow negative notch ($-0.45~\mu$A$\cdot$nm/V$^2$ at
$5.64$~eV) between the positive maxima at $5.4$ and $5.9$~eV: this is
the reversal interval of Fig.~\ref{fig:sweep}(b). 

The mechanism is general, and the pathway realizes it in three
regimes. At AA and SP the two bright channels are exactly degenerate and
displace in opposite directions, so they cancel at every frequency; the
zeros that symmetry imposes at these stackings are thus the limiting
case of the same compensation, complete there and only partial at AB
and BA. At AB and BA the near-compensation just described
produces the deep central reversal. At intermediate registries one
sign dominates the strong-absorption region and the obligatory
reversals retreat to the spectral edges; every stacking along the
sweep still crosses zero at least once, with one to three robust
crossings per stacking. This fixes the choice of
$5.6$~eV in Fig.~\ref{fig:sweep}(c): at that energy the curve reads
zero at all five high-symmetry stackings and close to its maxima at
the intermediate ones. 

\subsection{Shift vector, band polarizations, and the
oscillator-weight sum rule}\label{sec:shiftvector}

\begin{figure}[tb]
\includegraphics[width=\columnwidth]{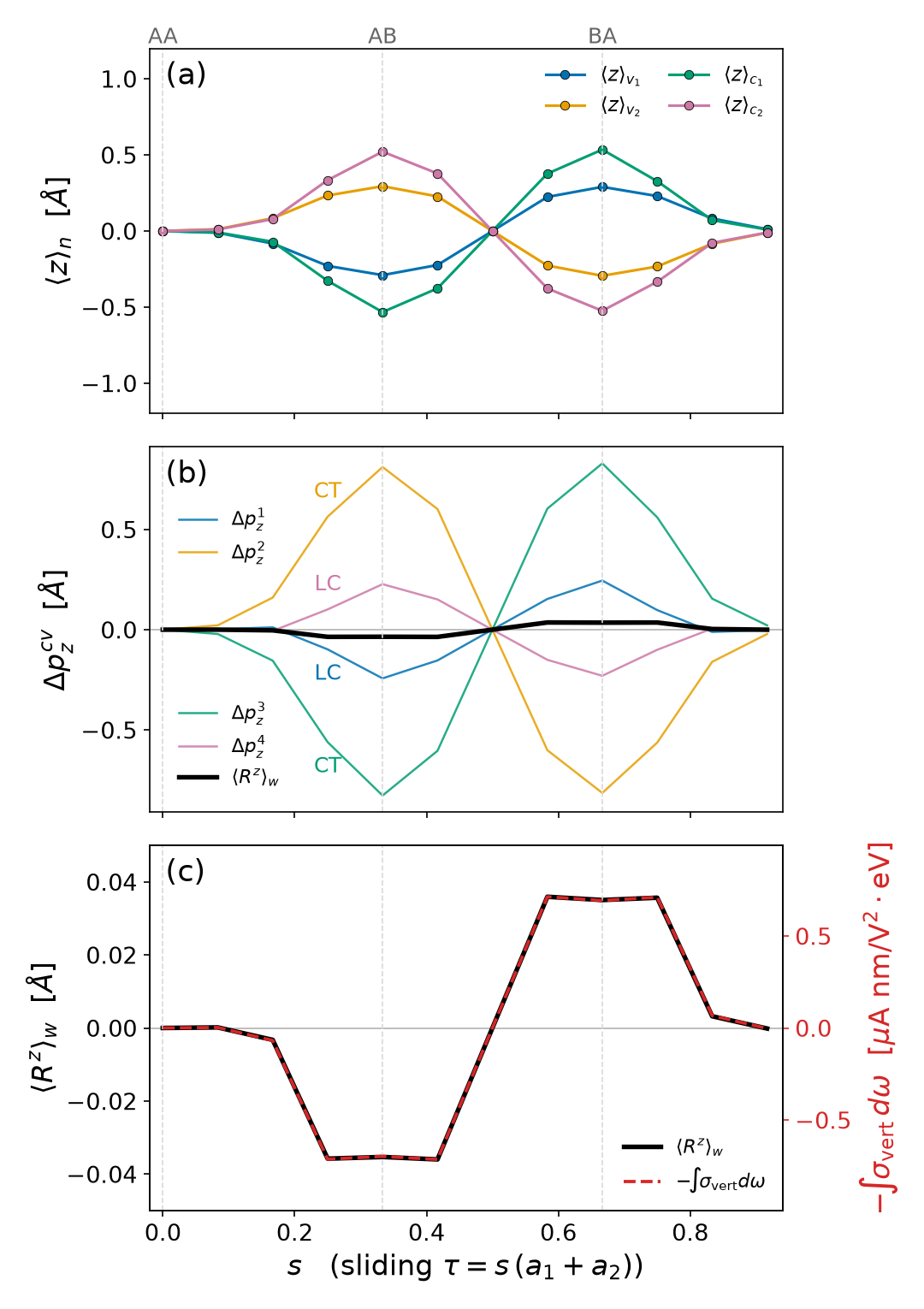}
\caption{\label{fig:shiftvector}Band-resolved anatomy of the vertical
response along the bilayer pathway. (a)~Vertical band
centroids $\langle z\rangle_n$ of the four $\pi$ bands, labeled in
ascending energy ($v_1$ the deepest valence, $c_1$ the lowest
conduction band; pairs in panel (b) follow the same convention,
$1\!-\!4 = v_1c_1, v_1c_2, v_2c_1, v_2c_2$). (b)~Unweighted band-centroid differences
$\Delta p_z^{cv} = \langle z\rangle_c - \langle z\rangle_v$ (thin
lines), labelled by channel family (CT, charge transfer; LC, layer
conserving; Sec.~\ref{sec:channels}), against the oscillator-weighted
shift vector $\langle R^z\rangle_w$ (black line, $\sim$20 times
smaller). (c)~Zoom on $\langle R^z\rangle_w$, against the integrated response
(dashed, right axis); the two axes differ by the constant factor
$(\pi e^3/\hbar^2)\,S_{\rm tot}$.}
\end{figure}

The vertical channel of a slab makes the formal link between the
shift vector and the polarization exact. By an identity of Fregoso,
Morimoto, and Moore \cite{Fregoso2017}, the BZ integral of the shift
vector equals the polarization difference between the two bands of
the transition. For the $z$ component of a slab the phase-derivative
term of $R$ is absent (Sec.~\ref{sec:model}), so the identity carries
no winding ambiguity and reads simply $\overline{R^z_{cv}} =
\langle z\rangle_c - \langle z\rangle_v$ by Eq.~\eqref{eq:rz}, with
the band centroids of Fig.~\ref{fig:shiftvector}(a).

If that identity controlled the response, the shift current would be
a polarization readout, precisely the assumption of
Ref.~\cite{NbOBr2}. It does not, because the response also weighs each transition by its optical strength [Fig.~\ref{fig:shiftvector}(b,c)]. The exact pair decomposition
$\int\!\sigma_{\rm vert}\,d\omega = (\pi e^3/\hbar^2)\sum_{cv}\int
R^z_{cv} w_{cv}$, with $w_{cv} = \tfrac12(|r^x_{cv}|^2+|r^y_{cv}|^2)$
the in-plane optical transition probability of the pair, reproduces
the spectra. Fregoso, Morimoto, and Moore proposed, on the basis of this identity, a quantitative relationship between the polarization differences of the bands and the zone-averaged shift current, so that one could be estimated from the other \cite{Fregoso2017}. That relationship amounts to treating the optical weight as uniform over the zone: the pair integral then factorizes into the product $\Delta p^{cv}_z \times S_{cv}$, where $\Delta p^{cv}_z \equiv
\overline{R^z_{cv}}$ is the unweighted zone average of the shift vector
of the pair (the band-centroid difference of the identity) and $S_{cv}
= \int_{\rm BZ}[d^2k]\, w_{cv}(\bm{k})$ is its total oscillator weight.
This factorized estimate fails: for the two charge-transfer pairs that
carry the response (Sec.~\ref{sec:channels}) it deviates from the exact
pair integral $\int R^z_{cv} w_{cv}$ by $15$--$46\%$ at the AB registry
and at the two neighboring slidings of the sweep. The reason is that
the shift vector and the oscillator strength are strongly
anticorrelated point by point in the zone: the strongest absorbers
displace charge the least (Sec.~\ref{sec:channels}), so the product of
the two averages does not reproduce the average of the product. 

The exact decomposition can nevertheless be condensed into a single
number. The total oscillator weight of the $\pi$ manifold, $S_{\rm
tot} = \sum_{cv} S_{cv}$, is essentially independent of the sliding:
it is set by the intralayer B--N transition dipole, with an interlayer correction of order $(t_\perp/E_g)^2$, $t_\perp$ being the interlayer hopping and $E_g$ the band gap (the integrated in-plane weight
varies by a few percent along every pathway computed here,
Sec.~\ref{sec:discussion}). Writing the integrated response as
$\int\!\sigma_{\rm vert}\,d\omega = (\pi e^3/\hbar^2)\,S_{\rm
tot}\,\langle R^z\rangle_w$ therefore places all of its sliding
dependence in the weighted shift vector
\begin{equation}
\langle R^z\rangle_w = \frac{1}{S_{\rm tot}}\sum_{cv}\int_{\rm
BZ}[d^2k]\, w_{cv}(\bm{k})\,R^z_{cv}(\bm{k}),
\label{eq:rw}
\end{equation}
the average of the shift vector over all transitions, each weighted by
its optical transition probability. The weights are essential. Without
them, the four unweighted averages $\Delta p^{cv}_z$ cancel in
near-opposite pairs [Fig.~\ref{fig:shiftvector}(b)] and their sum stays
below $0.01$~\AA{} at every sliding; with them, $\langle R^z\rangle_w$
is finite and retraces the integrated spectra along the whole pathway
[Fig.~\ref{fig:shiftvector}(c)]. The two curves of that panel are
computed independently from the same $\pi$ TBW Hamiltonian: $\langle
R^z\rangle_w$ from the pair decomposition and the integral from the
full shift-current spectra. 

The same centroids settle what the polarization itself is made of, a
question the sliding-ferroelectricity literature leaves open
\cite{Fan2023}: interlayer charge transfer \cite{Li2017,Ferreira2021}
or distortion of the $p_z$ clouds \cite{WangGui2023}? Within the
$\pi$ manifold the two separate exactly, and the answer is simple: the
layer transfer provides essentially all of $P^\pi$, the orbital dipole
a small correction, $0.1$--$0.2$~pC/m. What is not simple is the scale
[Fig.~\ref{fig:sweep}(a)]: along the entire pathway the
$\pi$-manifold polarization is proportional to the total,
\begin{equation}
P^{\pi}_z(s) \approx 3.8\, P_z(s),
\label{eq:screening}
\end{equation}
so the $\sigma$ manifold uniformly screens about three quarters of the
$\pi$ dipole. The uniformity itself is a bilayer feature: the
multilayers will break it (Sec.~\ref{sec:trilayer}). Single-channel
$\pi$ theories of stacking ferroelectricity \cite{Zhou2024,Yu2023} are
accurate for the channel they contain and miss the compensating ones;
the $\pi$-only model of Ref.~\cite{Yu2023} yields $P_z \simeq
2.2$~pC/m for the AB bilayer, matching our $\pi$-manifold polarization
($+2.5$) rather than the total ($+0.74$).

Eq.~\eqref{eq:screening} reflects internal dielectric screening. The $\pi$
interlayer transfer sets up a dipole, and with it an internal field.
The $\sigma$ orbitals polarize against that field; they comprise six of
the eight valence bands. The observed polarization is then $P^\pi$ divided by an effective transverse dielectric constant. The measured factor $3.8\pm0.2$ is, within its error bar, the out-of-plane dielectric constant of hBN, $\varepsilon_\perp\simeq 3.3$--$3.8$
\cite{Laturia2018}.


The shift current cannot see any of that $\sigma$ compensation, for
three independent reasons. First, every optical transition in
the 4.5--7~eV window is $\pi\to\pi^*$; the $\sigma\to\sigma^*$
continuum lies far above it. Second, within each
layer the $\sigma$ states are even and the $\pi$ states odd under the
layer mirror, so the in-plane dipole between the two sectors
vanishes. Both tensor components studied here absorb through in-plane
dipole matrix elements, and stacking breaks the layer mirror only
weakly, so $\sigma\to\pi^*$ transitions are dark to the same weak
order. Third, what remains available to the
$\sigma$ valence states is the role of virtual intermediates, entering
through matrix elements divided by energy separations of 5--10~eV.
That contribution is measured in Sec.~\ref{sec:bands}: a few-percent
effect.

The three reasons share one criterion, that of perturbation theory: a
remote state contributes in proportion to the square of its coupling
to the active $\pi$ states divided by its energy separation from them,
$|V|^2/\Delta E$, whatever the orbital character of that state. In
narrow-gap
$\pi$-electron multilayers such as graphene stacks, the ``other''
$\pi$ bands sit one interlayer hopping ($\sim$0.4~eV) away from the
active ones, and virtual channels are an order-one effect
\cite{PerezRiffo2026}; here the $\sigma$ sector fails the criterion
twice over, by weak coupling and by large separation. The asymmetry is therefore structural: the ferroelectric
order parameter is a screened observable; the photocurrent is unscreened.

\subsection{Charge-transfer versus layer-conserving channels}
\label{sec:channels}

\begin{figure*}[tb]
\includegraphics[width=0.92\textwidth]{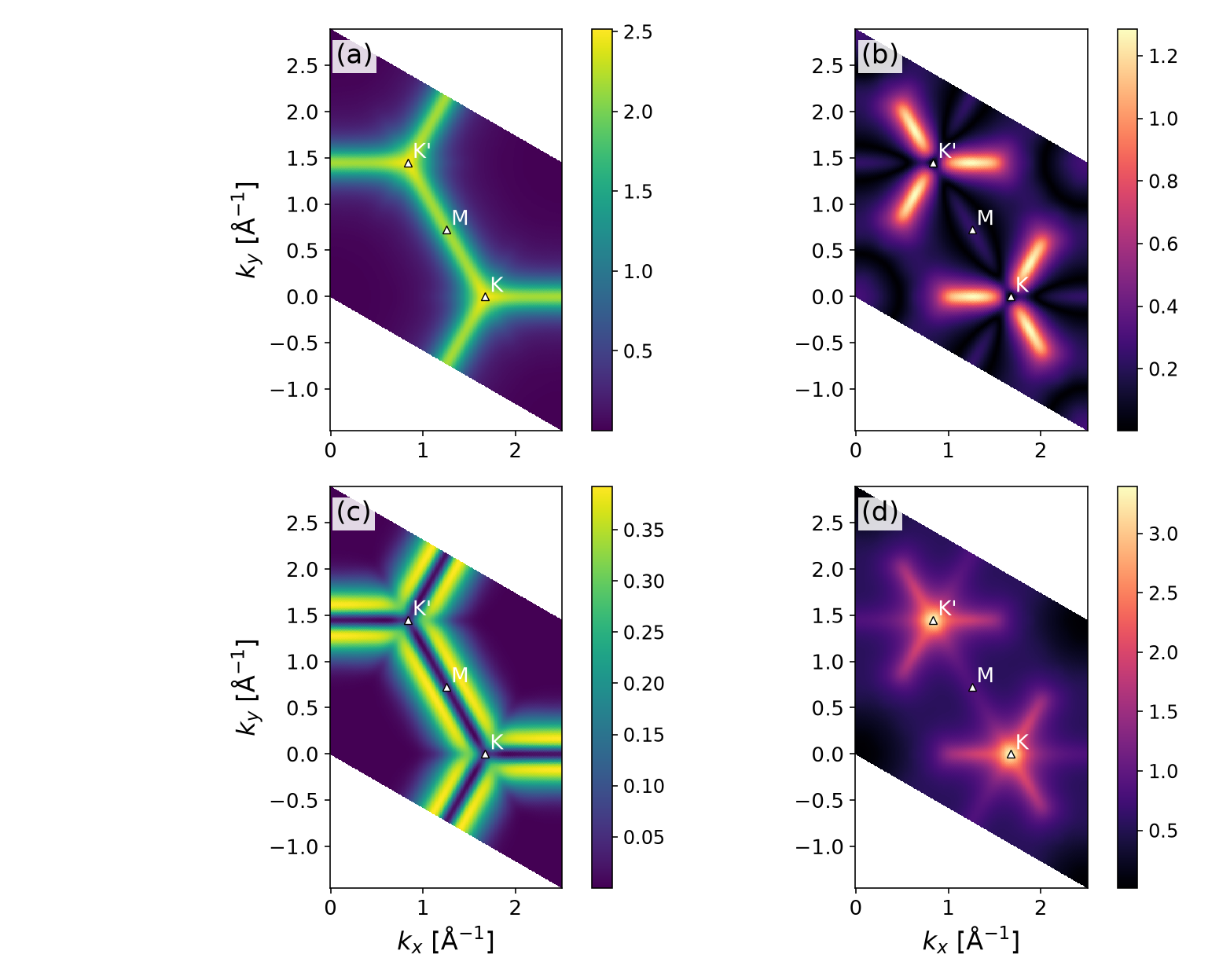}
\caption{\label{fig:channels}The weight--displacement anticorrelation
behind the vertical shift current (bilayer AB, $\pi$ TBW Hamiltonian,
$240^2$ mesh). Top row, the layer-conserving (bright) channels:
(a)~in-plane optical weight $w$ and (b)~vertical displacement
$|\Delta z|$. Bottom row, the charge-transfer channels: (c)~$w$ and
(d)~$|\Delta z|$. Within each row the two maps are complementary, and
between rows the contrast inverts. At K the bright family reaches the
largest weight of the whole zone, $w = 2.5$~\AA$^2$, while displacing
only $0.01$~\AA; the charge-transfer family displaces the full
interlayer distance, $3.4$~\AA, with a weight $130$ times smaller.
The bright ridges along K--M--K$'$ build the spectral structure; the
charge-transfer stars at K carry the broadband, domain-sign response.}
\end{figure*}

Now we analyze which transitions are responsible for the response. We have to consider two facts. First, the vertical band centroid measures nothing but the layer
weight of the state. The layers act as two rigid sheets: placing the
upper- and lower-layer weights of the wavefunction, $\rho^{u}_n$ and
$\rho^{l}_n$, at the sheet planes $z = \pm d/2$ reproduces the
centroid, $\langle z\rangle_n(\bm{k}) = (d/2)\,[\rho^{u}_n -
\rho^{l}_n]$, to within $0.1$~\AA{} across the zone and along the
pathway. Second, the
optical matrix element is intralayer: the in-plane velocity cannot
move an electron between layers. Together they split the four $\pi$
channels at every $\bm{k}$ into the two families of
Fig.~\ref{fig:channels}: layer-conserving (LC) channels, optically bright
but barely displacing [Fig.~\ref{fig:channels}(a,b)], and
charge-transfer (CT) channels, displacing by nearly the full interlayer
distance where the doublets are layer-pure, but dipole-weak
[Fig.~\ref{fig:channels}(c,d)]. These are the two sides of the
anticorrelation of Sec.~\ref{sec:shiftvector}. At the K point, the
brightest transitions of the whole zone displace the charge by only
$0.01$~\AA, while those displacing the full interlayer distance
absorb $130$ times less.

Which family carries the current then follows by elimination. The two
bright channels displace in opposite directions with nearly equal
weights, so the trace rule of Sec.~\ref{sec:bilayer} cancels them
almost completely at every $\bm{k}$. The two CT channels
also displace in opposite directions, but their weights are not
equal: the polar B--N contact is one-sided, hybridizing the
conduction sector of the layer below it with the valence sector of
the layer above it, and it therefore feeds one crossed transition
only (at K, weights $0.038$ against less than $10^{-3}$~\AA$^2$).
That channel is the net vertical current. At AB it is $1.2$ times
the net response; the LC family opposes the excess. Figure~\ref{fig:shiftvector}(b,c) shows the 
elimination process: the two families cancel in
near-opposite pairs, and the small weighted residue that survives is
$\langle R^z\rangle_w$ itself. 

Both statements are robust along the pathway: at the intermediate
slidings the CT family still supplies $108$--$120\%$ of the net
response, and the internal cancellation of the LC family stays at
$94$--$95\%$, its residue opposing only the excess. What reorganizes is the $k$-space texture: the
layer-pure locus deforms from wells at K into ridges connecting
K--M--K$'$, and the integrand concentrates along them, about three
times as intense as at AB. Hence the monochromatic response peaks at
the intermediate registries.

From the analysis above we can now trace each rule of this article
to its mechanism. Broadband illumination integrates the spectrum and
therefore reads the fed CT channel: its sign is the
orientation of the polar contact, it adds per interface, and the
result is a binary, domain-sign step. Monochromatic illumination
resolves the bright-pair interference, which is registry- and
frequency-sensitive by construction. Either the
photon supplies the displacement and hybridization lends the dipole
(CT), or the photon supplies the dipole and
hybridization lends the displacement (LC). In both
cases the vertical shift current is an interlayer-hybridization observable, which is also why its absolute magnitude, unlike its
structure, is controlled by the remote-band channels of
Sec.~\ref{sec:bands}. In the bilayer it only selects the fed channel; in a stack with more than one interface, an interface acquires a position, and which sector of the optically active states
it perturbs depends on its location. That is the mechanism, we dissect in Sec.~\ref{sec:trilayer}, behind equal-polarization stackings with opposite photocurrents.

\section{The nine trilayer stackings}
\label{sec:trilayer}

\begin{figure}[tb]
\includegraphics[width=\columnwidth]{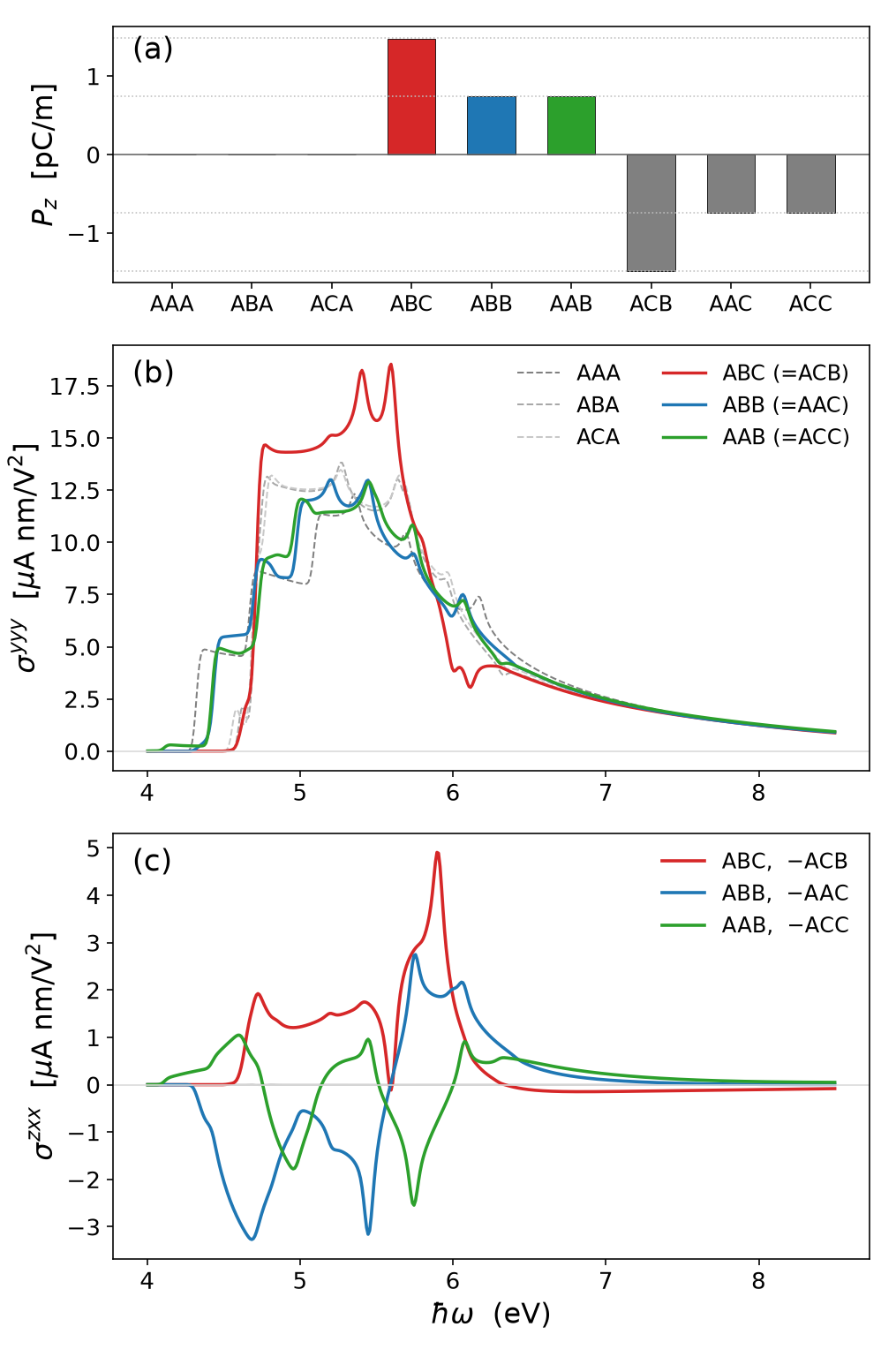}
\caption{\label{fig:spectra}Polarization and shift current across the
nine trilayer stackings. (a)~First-principles $P_z$ per stacking: the
steps of interface additivity, with guides at $\pm p_0$ and $\pm 2p_0$
($p_0 = 0.74$~pC/m). (b)~In-plane $\sigma^{yyy}$ ($\pi$ TBW
Hamiltonians): substantial for every stacking (nonpolar $D_{3h}$
dashed, about $0.7$ of the polar ones). (c)~Vertical $\sigma^{zxx}$,
one curve per mirror pair (the partner is its exact negative):
$D_{3h}$ stackings vanish, ABC (two aligned AB interfaces) is the
largest, and the equal-polarization pair ABB/AAB differ in sign.}
\end{figure}

With three layers the landscape acquires structure: we can count the interfaces and see how they are stacked. Across the nine
stackings, the first-principles dipole obeys interface additivity
[Fig.~\ref{fig:spectra}(a)]: each AB-type (AC, BC) interface contributes
$\pm p_0$ ($p_0 = 0.74$~pC/m from the QE dipole), while AA-type
interfaces, locally nonpolar, do not contribute. The three $D_{3h}$
stackings therefore give $P_z = 0$; the six single-AB-interface
stackings give exactly $\pm p_0$; and ABC/ACB give $\pm 1.48$, twice $p_0$. 

The nine stackings also test the proportionality of
Eq.~\eqref{eq:screening}; we elaborate on why it is a bilayer feature. The orbital-$\pi$
fraction of the polarization is strongly stacking-dependent: ABB's
polarization is $\pi$-dominated ($P^\pi \simeq P_z$), while AAB's,
with the identical total, carries an almost vanishing $\pi$ component
and is essentially $\sigma$-borne.

The origin lies in how the polar B-N contact is assembled: B-N-N in
ABB, B-B-N in AAB (bottom to top). In B-N-N the direct interlayer contact that joins the polar
interface is the nitrogen pair, coupled by $t_{NN}$; in B-B-N it is
the boron pair, coupled by $t_{BB}$. The two hoppings are unequal
(Sec.~\ref{sec:tb}). Intuitively, the eclipsed pair might become a path for the charge that the polar contact transfers. In
B-B-N that path runs through B-B: two nearly empty $p_z$ orbitals, on
the conduction side of the gap, coupled by the largest interlayer
hopping of the parametrization; that bridge lets
the transferred charge flow back and cancels the $\pi$ dipole almost
entirely. In B-N-N the path runs through N-N: two full, deeply bound
$p_z$ orbitals coupled by the smallest hopping; a stiff bridge that
transmits little and leaves the $\pi$ transfer intact. Quantitatively,
$P^{\pi}/P_z = +1.05$ for the B-N-N (N-N-B) family, ABB and AAC,
and $P^{\pi}/P_z \simeq -0.3$ for the B-B-N (N-B-B) family, AAB and
ACC. 

The in-plane response $\sigma^{yyy}$ is robust for all nine stackings
[Fig.~\ref{fig:spectra}(b)], including the three nonpolar $D_{3h}$
ones, and the polar stackings are larger only by a factor of about
$1.4$. An out-of-plane polarization is therefore not a requirement for
this channel: in-plane inversion breaking, boron against nitrogen, is
already sufficient (Sec.~\ref{sec:symm}). The mirror pairs behave as
the switching rule requires: $\sigma^{yyy}$ is identical within each
pair (an even number of $z$ indices) to a fraction of a percent,
including the crossed pairings ABB/AAC and AAB/ACC of
Table~\ref{tab:symm}.

We now turn to the vertical response. The three $D_{3h}$ stackings
vanish identically. The ABC stacking, with two aligned AB interfaces, has the largest $\sigma^{zxx}$ of all trilayers [Fig.~\ref{fig:spectra}(c)]
($\simeq 4.9~\mu$A$\cdot$nm/V$^2$, close to twice the bilayer value). Its spectrum remains single-signed over most of the energy range,
and the ACB spectrum is its exact negative. This largest response is
precisely the one the TB Hamiltonian predicts to vanish, the
qualitative failure anticipated in Sec.~\ref{sec:tb}. 

The ABB and AAB stackings have smaller responses, and their spectra
change sign at least once.
However, the correspondence
with the polarization ends there. The shift current weighs each transition by its vertical jump
[Eq.~\eqref{eq:rz}] and is therefore sensitive to the interface's
height in the stack.

Table~\ref{tab:decoupling} is the resulting counterexample. ABB and AAB
carry a single AB-type interface each, at the bottom of one slab and at
the top of the other, and their first-principles polarizations are
identical ($+0.74$~pC/m). Yet their vertical shift conductivities
have opposite signs for a large part of their spectra, and opposite frequency integrals
($-0.58$ versus $+0.25$).
(Why moving the interface flips the sign has a microscopic answer,
the one-sided contact of Sec.~\ref{sec:channels}.)

The failure of a polarization reading is not a matter of the chosen
frequency. Table~\ref{tab:decoupling} samples the two dominant peaks
of the set. At $4.7$~eV the strongest response is ABB's, negative
against its positive polarization; at $5.7$~eV ABC only matches ABB
despite twice the polarization, and AAB opposes both. For instance, there is no single energy in the $\pi$ window where
the three responses share a sign, with ABC the largest
[Fig.~\ref{fig:spectra}(c)], and the frequency integrals do no
better. No measurement of the vertical photocurrent
could rank, or even sign, these polarizations: what $\sigma^{zxx}$
encodes is the vertical registry of the polar interfaces, not the
polarization they sum to. This is the minimal, material-specific
realization of the anomalous behavior reported in CuInP$_2$S$_6$
\cite{Bai2024}.

\begin{table}[tb]
\caption{\label{tab:decoupling}Vertical shift conductivity (sheet values,
$\mu$A$\cdot$nm/V$^2$, $\pi$ TBW Hamiltonians) for three polar
trilayers at the two dominant peaks of the set, its frequency integral, and the first-principles $P_z$
(QE dipole). Equal polarization, as in the pair ABB, AAB, does not imply
equal or even same-sign vertical responses.}
\begin{tabular}{lcccc}
\toprule
 & $P_z^{\rm QE}$ (pC/m) & $\sigma^{zxx}(4.7\,\mathrm{eV})$ &
 $\sigma^{zxx}(5.7\,\mathrm{eV})$ & $\int\!\sigma^{zxx}d\omega$\\
\midrule
ABC & $+1.48$ & $+1.81$ & $+2.84$ & $+2.39$\\
ABB & $+0.74$ & $-3.20$ & $+2.73$ & $-0.58$\\
AAB & $+0.74$ & $+0.50$ & $-2.55$ & $+0.25$\\
\bottomrule
\end{tabular}
\end{table}

\section{The tetralayer sliding pathway}
\label{sec:tetralayer}

\begin{figure}[tb]
\includegraphics[width=\columnwidth]{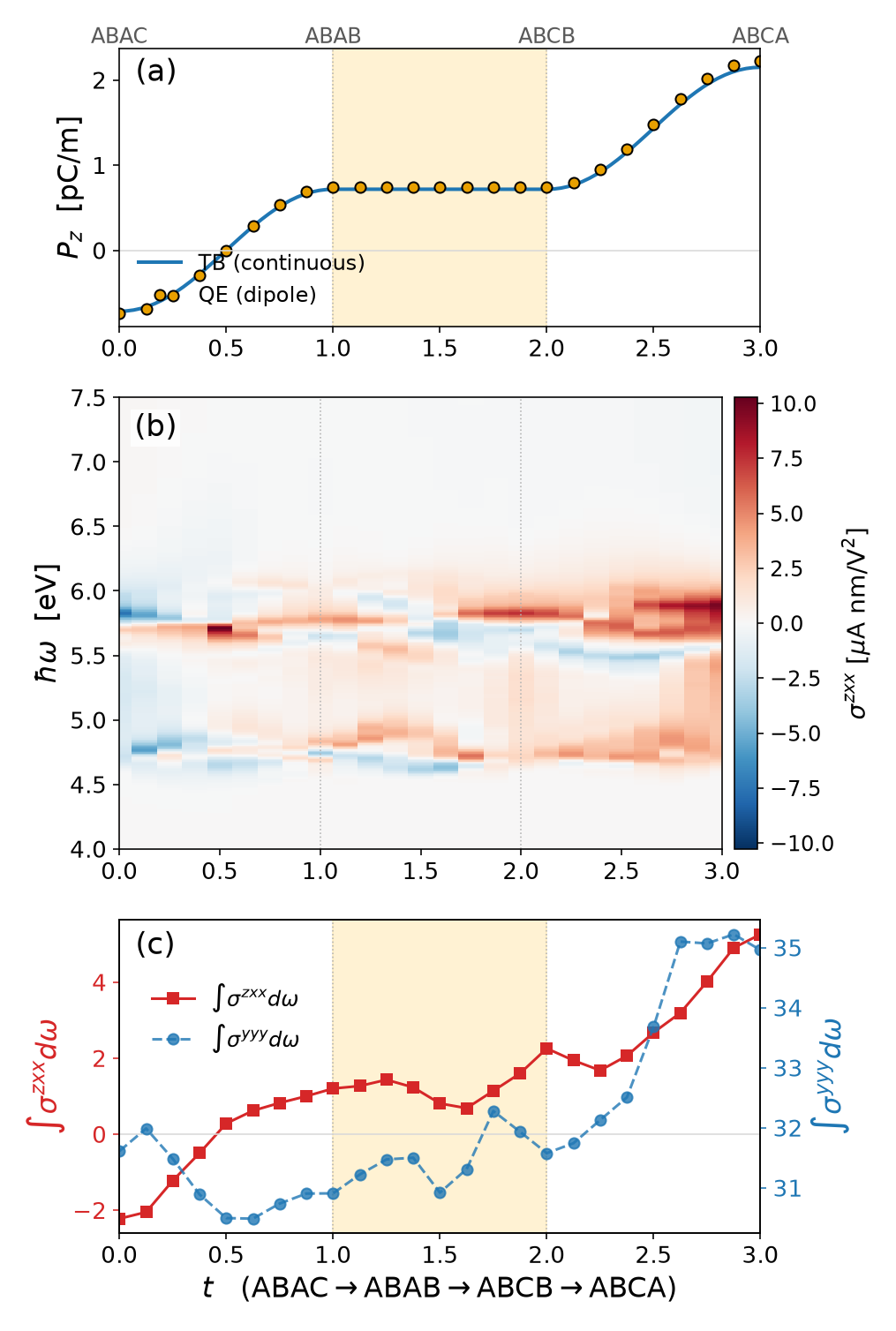}
\caption{\label{fig:plateau}First-principles sweep of the tetralayer
pathway ABAC$\to$ABAB$\to$ABCB$\to$ABCA (25 slidings, eight-band Wannier
models). (a)~$P_z$: QE dipole (points) on the TB curve (line);
on the shaded plateau the first-principles polarization is constant to the
last printed digit of the dipole ($0.006$~pC/m). (b)~Spectral map
$\sigma^{zxx}(\omega,t)$: the response reshapes and changes sign
inside the plateau. (c)~Frequency-integrated responses: the vertical
integral varies threefold across the constant-$P_z$ plateau while the
in-plane integral stays rigid.}
\end{figure}

The tetralayer shows that decoupling is not incidental. 
Figure~\ref{fig:plateau} follows the low-barrier pathway
ABAC$\to$ABAB$\to$ABCB$\to$ABCA of Ref.~\cite{Vizcaya2025}, computed entirely from first principles at 25 slidings. The pathway coordinate $t \in [0,3]$ counts the successive slidings: each unit of $t$ displaces one layer by one bond length and takes the stack from one registry stacking to the next (ABAC at $t = 0$, ABAB at $t = 1$, ABCB at $t = 2$, ABCA at $t = 3$), and $\sigma^{zxx}(\omega,t)$ in Fig.~\ref{fig:plateau}(b) is the vertical spectrum at each point of the pathway. The middle segment is a polarization plateau. As one layer slides by a full bond length, $P_z$
stays between $0.737$ and $0.743$~pC/m. The plateau exists because the sliding layer trades one
AB-type interface for another, leaving the interface count, and hence the polarization, untouched, while the stacking changes continuously.

The vertical response, however, is not frozen: although always
positive in its integral, it ranges from $+0.68$ to
$+2.25~\mu$A$\,$nm/V$^2\,$eV, a factor of $3.3$. At
fixed $\hbar\omega = 5.6$~eV, $\sigma^{zxx}$ reverses sign ($+2.2 \to -1.7$). A monochromatic vertical response might reverse while the polarization remains constant.

The mirror-conjugate pair ABAC ($t=0$) and ABCB
($t=2$)
gives opposite $\int\!\sigma^{zxx}d\omega$ ($-2.23$ versus $+2.25$) with
identical $\int\!\sigma^{yyy}d\omega$ ($0.1\%$), following the switching rule. The
polarization obeys the counting rule along the way (ABAC $-0.743$;
plateau $+0.740$; ABCA $+2.224 = 3\times0.741$, the same $p_0$ as bilayer
and trilayer). However, the largest vertical peak of the
entire pathway ($\sigma^{zxx} = +10.3~\mu$A$\cdot$nm/V$^2$ at $t = 0.5$, $\hbar\omega = 5.70$~eV)
occurs precisely where $P_z$ crosses zero.
\section{The role of remote bands}\label{sec:bands}

Everything above rests on $\pi$ TBW Hamiltonians ($2N$ bands for
$N$ layers), so the natural objection would be the role of the virtual channels \cite{Chen2024,PerezRiffo2026}. The sum
over intermediate states in Eq.~\eqref{eq:gencov} is truncated to the
$\pi$ manifold, and the $\sigma$, $\sigma^*$ and nearly-free-electron (NFE) bands it omits carry large interlayer dipoles. How much do they
matter? We address this question now.

We performed a calculation enlarging the manifold from $2N$ to $5N$ Wannier functions per stack. They include the complete valence space (three $\sigma$
bonds and one $\pi$ per layer) plus the $\pi^*$ conduction band. Every
$\sigma$-valence virtual channel, and the full dipole structure of the occupied space, is therefore now contained in the model. We carried
this out for the bilayer AB, the trilayer ABC and the tetralayer ABCA,
comparing the $\pi$ and full-valence TBW Hamiltonians on identical frequency and $k$-grids. Although there are some deviations, they do not question the previous results. 


The enlargement stops at $5N$ for a structural reason. The $\sigma^*$ manifold of a thin slab cannot be added to any local-orbital model, because it does not exist
as a localized subspace. Between $\sim$2 and 7~eV above the gap the atomic projectability of the slab conduction states collapses to $0.03$--$0.34$; they hybridize irreversibly with the vacuum NFE states. Attempts to wannierize them either capture vacuum-centered functions or fail.

These non-wannierizable channels require caution. A single full-band calculation of the sliding bilayer reports the vertical component
\cite{Kaner2022}; the other first-principles study reports in-plane
components only \cite{Agarwal2025}. The sheet-converted peaks of that
calculation ($\sigma_{\rm 3D}$ times its effective bilayer thickness,
$6.58$~\AA) exceed the corresponding $\pi$ TBW peaks by a factor
$\sim$3 in plane ($\approx 33$ versus $10.4~\mu$A$\cdot$nm/V$^2$)
and $\sim$6--7 in the vertical channel ($\approx 10$ versus $1.5$). Most of the
vertical spectral weight thus originates in the remote conduction bands, the same states behind the vertical failure of the minimal TB parametrizations (Sec.~\ref{sec:trilayer}). There is no
contradiction with the convergence just found: the enlargement exhausted the channels a local model can hold, and the weight that is still missing lives precisely in the states no local model can contain.

The information content of the article survives this rescaling.
Mirror parity, the zeros of the nonpolar stackings and the switching
rules are fixed by symmetry and hold. The
stacking contrasts compare slabs of identical composition, so an enhancement common to the set drops out of the ratios, although only a full-band campaign could certify this cancellation. Absolute magnitudes, in contrast, must be read as lower bounds, by roughly the
factors quoted above. The same caution extends beyond the
independent-particle approximation: excitons enhance and red-shift the
hBN response by close to an order of magnitude \cite{Huang2023}.


\section{Discussion}
\label{sec:discussion}

The relation we set out to find, between the polarization and the shift
current of a sliding ferroelectric, turns out to be exactly one rule. It ties the sign of the integrated vertical response to the ferroelectric state, in every system we
computed. Nothing else survives. Magnitudes decouple (trilayer ABC),
signs decouple at fixed $|P_z|$ (ABB versus AAB), spectra reorganize at
frozen polarization (the tetralayer plateau), and the strongest responses
live where the polarization vanishes (intermediate bilayer slidings; the
$t=0.5$ crossing of the tetralayer).

The mechanism is a separation of
functionals. $P_z$ counts polar interfaces, an occupied-state property built from strongly compensating $\pi$ and $\sigma$ contributions whose ratio is stacking-dependent; $\sigma_{\rm vert}$ weighs
$\pi$-manifold transitions by their vertical displacement, and thereby reads interface positions. 

Both behaviors reported in the literature are thus understood. Correlation appears when an experiment
compares the two states of a switching cycle: that is the one
comparison where the switching rule forces $P_z$ and $\sigma_{\rm vert}$
to reverse together \cite{NbOBr2}. Decoupling appears under every other variation: the registry can move at fixed interface count, reversing the photocurrent at constant polarization \cite{Bai2024}, and added
layers grow the polarization by counting while the spectra reorganize
\cite{Delodovici2025,Xu2025}.

In practice, each channel of the response reads a different property. Monochromatic
photocurrent maps can read the registry: at fixed photon energy the vertical response (the bright-pair interference of
Sec.~\ref{sec:channels}) distinguishes stackings that the
polarization cannot, vanishing
at high-symmetry registries and peaking at intermediate ones. The same
contrast resolves domain walls and partial slidings in vertical
graphene/hBN/graphene devices \cite{Yasuda2021,Xu2025}. 

The monochromatic reading is a capability with a caveat. Away from
the symmetry-protected zeros and the mirror pairs, the value at a
fixed photon energy is set by the interference of transitions, which
is specific to the band structure of each material. Its exact
behavior is not known in general. Using it as a registry probe
requires a detailed study of how the response peaks evolve and of
their correspondence with the stackings. On the other hand, broadband
illumination reads the domain sign: the frequency-integrated vertical
response (carried by the charge-transfer channels of
Sec.~\ref{sec:channels}) is a step function of the ferroelectric state, insensitive to
$|P_z|$ and to registry within a plateau. That makes it a binary, optically read domain label. 

The in-plane channel is non-switchable, sliding-invariant in its integrated weight ($\pm3$--$5\%$ across every pathway computed here). Currents from opposite domains add rather than cancel.

A perpendicular displacement field might offer the decisive test of the decoupling. Within the TB Hamiltonian, a bias tuned so that the out-of-plane polarization reads exactly zero leaves the integrated vertical photocurrent finite, and in fact enhanced, at the AB registry. The reason is that the field conditions
that null the screened polarization are far from those that
null the resonant, charge-transfer-borne photocurrent. A systematic study of this field-tuned decoupling is left for future work.

None of this appears specific to hBN. The mechanism requires
only two structural ingredients, interlayer coupling dictated by the
registry and a sublattice asymmetry that breaks the electron--hole
equivalence, so the dichotomy between a counting polarization and a
position-reading photocurrent should extend to the
sliding-ferroelectric class. A minimal-model analysis that makes this
statement quantitative, together with the field-tuned decoupling
above, will be presented elsewhere.

A word of caution. Our spectra are independent-particle
calculations; excitons reshape and enhance the hBN response
\cite{Huang2023}, so peak positions and magnitudes are quasiparticle
baselines. The symmetry-protected statements, by contrast, survive
interactions unchanged. Lattice relaxation along pathways was neglected (rigid slides
at matched interlayer distance), controlled by the pathway-level
polarization validation. 

\section{Conclusions}\label{sec:conclusions}

We computed the polarization and the full shift-current tensor of hBN
multilayers side by side, from first principles: forty-six stackings,
two to four layers. The two quantities share their symmetry, the
mirror-dictated switching rule verified in every system, and nothing
else. The polarization counts polar interfaces: $P_z \simeq n_{\rm AB}\,p_0$, with $n_{\rm AB}$ the number of AB-type interfaces and the same $p_0 \simeq 0.74$~pC/m in bilayer, trilayer and tetralayer. The shift current reads where those interfaces sit.

For experiments, the message is direct. Broadband illumination reads the
polarization through its sign, as a binary domain label. Monochromatic
illumination could instead detect the stacking itself; that use
requires a detailed study of the optical transitions of each system.
The in-plane channel, sliding-invariant, provides the calibration.
The natural next step is a perpendicular displacement field, which
within the tight-binding Hamiltonian nulls the polarization while
leaving the photocurrent finite; that study is underway. The BPVE in
sliding ferroelectrics is a stacking spectroscopy.

\begin{acknowledgments}
E.S.M. acknowledges financial support from Proyecto Interno USM 2026
PI\_LIR\_26\_12.
\end{acknowledgments}

\section*{Data availability}
The numerical data underlying all figures, and the scripts that
generate them, are available from the corresponding author upon
reasonable request.

\section*{AI-use disclosure}
An AI assistant (Claude, Anthropic; Opus and Fable models) was used to
assist with the generation of scripts for the calculations and for
plotting the figures. The assistant was directed through interactive
prompting, and every AI-suggested script and edit was reviewed and
verified by the authors. All physical content, derivations, and
numerical results were produced by the authors, who take full
responsibility for the final text. No figure was generated or modified
by AI; all figures are plots of computed data produced by the authors'
scripts.

\bibliography{Bibliografia}

\end{document}